# FashionBERT: Text and Image Matching with Adaptive Loss for Cross-modal Retrieval

Dehong Gao, Linbo Jin, Ben Chen, Minghui Qiu, Peng Li, Yi Wei, Yi Hu and Hao Wang
Alibaba Group
Hangzhou City, Zhejiang Province, China
{dehong.gdh, yuyi.jlb, chenben.cb, minghui.qmh, jerry.lp, yi.weiy, erwin.huy, longran.wh}@alibaba-inc.com

## ABSTRACT

In this paper, we address the text and image matching in cross-modal retrieval of the fashion industry. Different from the matching in the general domain, the fashion matching is required to pay much more attention to the fine-grained information in the fashion images and texts. Pioneer approaches detect the region of interests (i.e., RoIs) from images and use the RoI embeddings as image representations. In general, RoIs tend to represent the "object-level" information in the fashion images, while fashion texts are prone to describe more detailed information, e.g. styles, attributes. RoIs are thus not fine-grained enough for fashion text and image matching. To this end, we propose FashionBERT, which leverages patches as image features. With the pre-trained BERT model as the backbone network, FashionBERT learns high level representations of texts and images. Meanwhile, we propose an adaptive loss to trade off multitask learning in the FashionBERT modeling. Two tasks (i.e., text and image matching and cross-modal retrieval) are incorporated to evaluate FashionBERT. On the public dataset, experiments demonstrate FashionBERT achieves significant improvements in performances than the baseline and state-of-the-art approaches. In practice, FashionBERT is applied in a concrete cross-modal retrieval application. We provide the detailed matching performance and inference efficiency analysis.

## 1 Introduction

Over the last decade, a great number of multimedia data (including image, video, audio, and text) have been emerged on The internet. To search important information from these multi-modal data efficiently, multimedia retrieval is becoming an essential technique and widely researched by world-wide researchers. Recently, it has been witnessed a soar increase of the research interest in cross-modal retrieval, which takes one type of data as the query and retrieves relevant data of another type. The pivot of cross-modal retrieval is to learn a meaningful cross-modal matching [40].

There exists a long research line in cross-modal matching, especially in text and image matching. The early approaches usually project visual and textual modal representations into a shared embedded subspace for the cross-modal similarity computation or fuse them to learn the matching scores, for example, the CCA-based approaches [14, 25, 44] and the VSE-based approaches [10, 11, 18, 41]. Very recently, the pre-training technique has been successfully applied in Compute Visual (CV) [1, 2] and Nature Language Processing (NLP) [8, 46]. Several researchers are inspired to adopt the pre-trained BERT model as the backbone network to learn the cross-modal information representation [19, 34]. The proposed approaches have achieved promising performances on several down-stream tasks, such as cross-modal retrieval [40], image captioning [1] and visual question answering [2]. However, these studies are centered on text and image matching of the general domain. In this paper, we focus on the text and image matching of the fashion industry[1], which is mainly referred to clothing, footwear, accessories, makeup and etc.

---

[1] https://en.wikipedia.org/wiki/Fashion

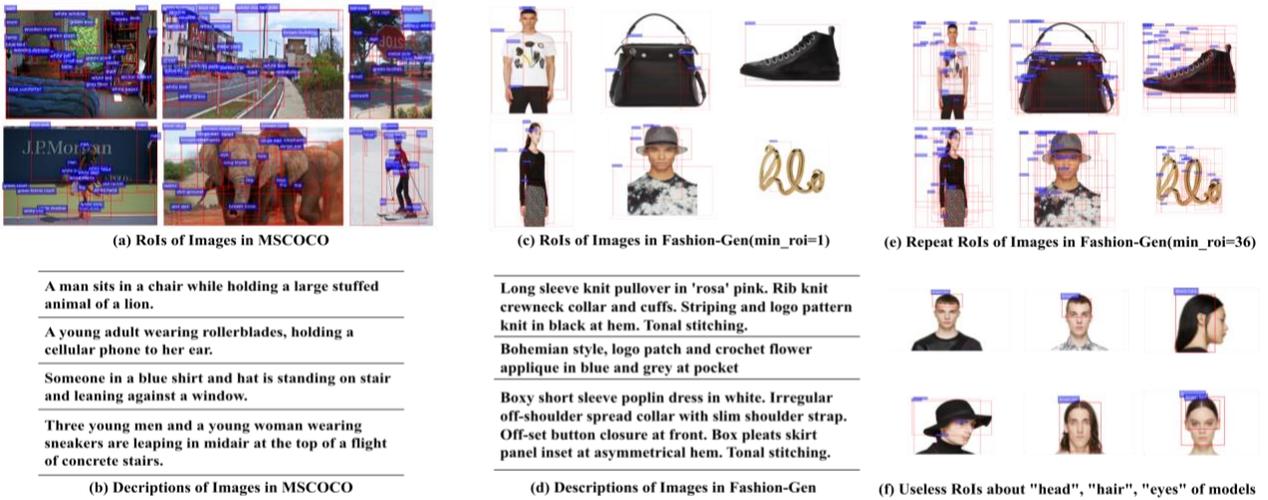

Figure 1: Comparison of text and image in the general and fashion domains. (a) and (b) are the RoIs and descriptions of MSCOCO Images from the general domain. (c) and (b) are the relatively-rare RoIs and fine-grained descriptions of Fashion-Gen Images from the fashion domain. (e) and (f) are large amount of the repeated and useless RoIs detected from fashion images.

The main challenge of these pioneer matching approaches is how to extract the semantic information from images, and integrate this information into the BERT model. All current approaches detect RoIs[2] (i.e., Region of Interest [13]) from images as seen in Figure 1(a) and treat these RoIs as "image tokens". But this RoI method does not work well in the fashion domain since relatively-rare RoIs can be detected from fashion images. As seen in Figure 1(b), we show the detected RoIs of Fashion-Gen images of different categories, where the minimum number of detected RoIs is set to one from an image. We found on average 19.8 RoIs can be detected from one MSCOCO[3] image, but only 6.4 can be detected from one Fashion-Gen[4] image. This is because in general a fashion image contains only one or two objects (e.g., a coat and/or a pant) with a flat background. We can set the minimum RoI number to detect, but under this setting lots of detected RoIs are repeated since they only focus on the same object(s) as seen in Figure 1(e). These repeated RoIs will produce similar features and contribute little to the later modeling. Meanwhile, we find some RoIs from fashion images are useless for text and image matching, for example, RoIs about the body parts (head, hair, hands etc.) of the models in fashion images as seen in Figure 1(f). These RoIs are irrelated to the fashion products and cannot build connection with the descriptions. On the contrary, most of the fashion texts describe the fine-grained information about the products (e.g., "crew neck", "off-shoulder", "high collar"). Occasionally, some of descriptions contain abstract styles, e.g., "artsy" and "bohemian" as seen in Figure 1(d). The RoIs in fashion images can indicate main fashion object(s), but fail to distinguish these fine-grained attributes or styles. Thus, it is more difficult for fashion text and image matching with such "object-level" RoIs and fine-grained descriptions.

In this paper, we propose FashionBERT to solve the above problems. Inspired by the selfie idea [38], we first introduce the patch method to extract image tokens. Each fashion image is split to small patches with the same pixels and we assume these patches as image tokens. The patches show more rawer pixel information, and thus contain more detained information compared with object-level RoIs. Besides, the split patches are non-repeated and ordered in nature, which are well suitable as the sequence inputs of the BERT model. The training procedure of FashionBERT is a standard multitask learning procedure (i.e., Masked Language Modeling, Masked Patch Modeling and Text&Image Alignment, which will be depicted in the later section). We propose an adaptive algorithm to balance the learning of each task. The adaptive algorithm treats the determination of loss weights of each task as a new optimal problem and will estimate the loss weights in each batch step.

We evaluate FashionBERT with two tasks, Text&Image alignment classification and cross-modal retrieval (including Image-to-Text and Text-to-Image retrieval). Experiments are conducted on the public fashion product dataset (Fashion-Gen). The results show that FashionBERT significantly outperforms the SOTA and other pioneer approaches. We also apply FashionBERT in our E-commercial website. The main contributions of this paper are summarized as follows:

1) We show the difficulties of text and image matching in the fashion domain and propose FashionBERT to address these issues.

2) We present the patch method to extract image tokens, and the adaptive algorithm to balance the multitask learning of FashionBERT. The patch method and the adaptive algorithm are task-agnostic, which can be directly applied in others tasks.

---

[2] https://github.com/peteranderson80/bottom-up-attention
[3] http://cocodataset.org
[4] https://fashion-gen.com/

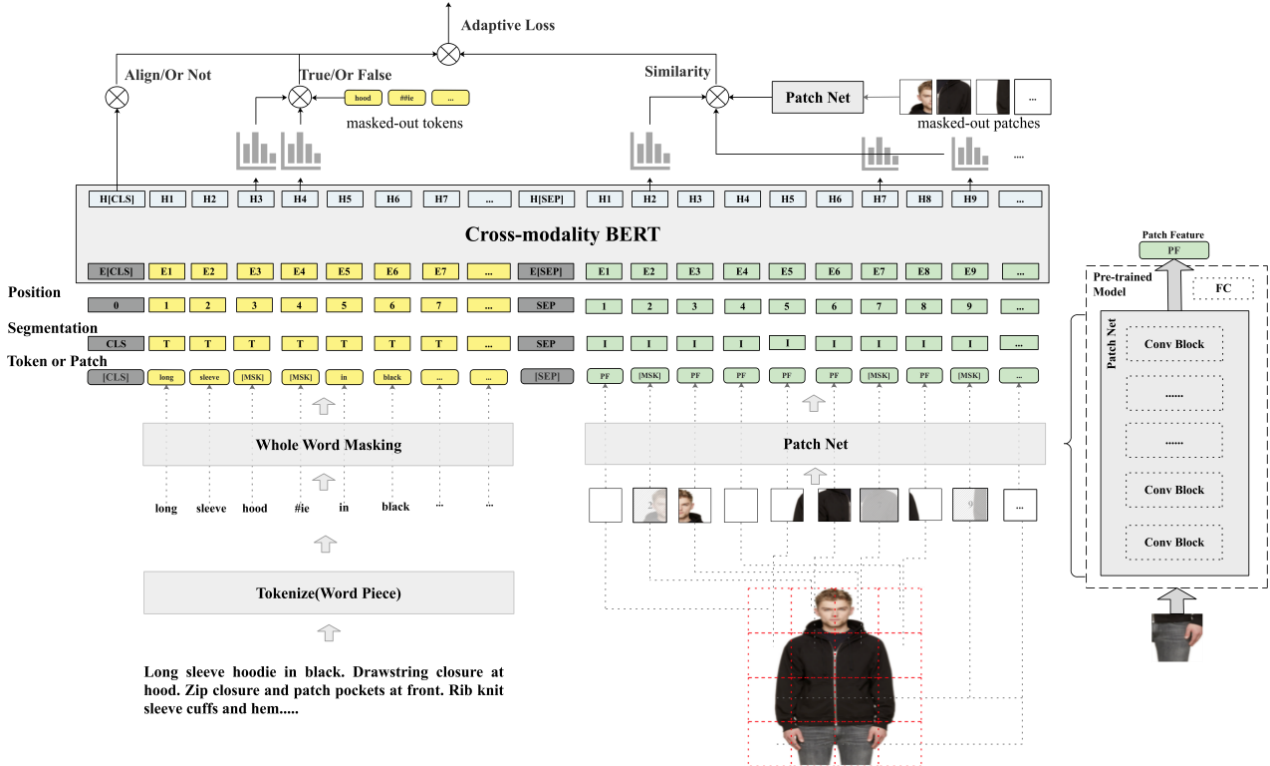

**Figure 2: our FashionBERT framework for text and image matching. We cut each fashion image into patches and treat these patches as "image tokens". After the interaction of text tokens and image patches in BERT, three tasks with adaptive loss weights are proposed to train the entire FashionBERT model.**

3) We extensively experiment FashionBERT on the public dataset. Experiments show the powerful ability of FashionBERT in text and image matching of the fashion domain.

4) FashionBERT currently has been applied in practice. We present the concrete application of FashionBERT in cross-modal retrieval. Meanwhile, we analyze both matching performances and inference efficiencies in detail.

## 2. Methodology

In this section, we will briefly revisit the BERT language model and then describe how we extract the image features and how FashionBERT jointly models the image and text data.

### 2.1 BERT

The BERT model introduced by [8] is an attention-based bidirectional language model. Taking tokens (i.e., word pieces) as inputs, BERT processes the embeddings of tokens with a multi-layer Transformer encoder [39]. When pre-trained on a large language corpus, BERT has proven to be very effective for transfer learning in variants of natural language processing tasks.

The original BERT model focuses on encoding of the single-modality text data. In the cross-modal scenario, the extended BERT model takes multi-modality data as input and allows them to interact within the Transformer blocks.

### 2.2 FashionBERT

The overview of FashionBERT is illustrated in Figure 2. It is composed of four parts, text representation, image representation, matching backbone and FashionBERT training with adaptive loss.

**Text Representation**: Similar to [8], the input text is first tokenized into a token sequence according to WordPieces [42]. The same BERT vocabulary is adopted in our FashionBERT model. We use the standard BERT pre-process method to process the input text. Finally, the sum of the word-piece embedding, position embedding and segmentation embedding is regarded as the text representation. The segmentation (i.e., "T" and "I" in Figure 2) is used to differentiate text and image inputs.

**Image Representation**: Different from the RoI method, we cut each image into patches with the same pixels as illustrated in Figure 2. We regard each patch as an "image token". For each patch, the outputs of the patch network are regarded as the patch features. It is possible to select any pre-trained image model, (e.g., InceptionV3 [36] and ResNeXt-101 [43]) as the backbone of the patch network. These patches are ordered in nature. The spatial positions of the patches are used in the position embedding. The sum of the patch features, the position embedding and segmentation embedding are regarded as patch representations.

**Matching Backbone**: The concatenation of the text token sequence and image patch sequence consists of the FashionBERT inputs. Similar to BERT, the special token [CLS] and separate token [SEP] are added in the first position and between the text token sequence and the image patch sequence, respectively.

The pre-trained standard BERT is adopted as the matching backbone network of FashionBERT. The information of text tokens and image patches thus interact freely in multiple self-attention layers. FashionBERT outputs the final representations of each token or patch.

**FashionBERT Training with Adaptive Loss**: We exploit three tasks to train FashionBERT.

Masked Language Modeling (MLM): This task is very similar to the MLM task utilized in BERT pre-training. We apply the Whole Word Masking (WWM) strategy to mask out all the text tokens corresponding to a word at once [6]. For example, in Figure 2, "long sleeve hoodie in black" is masked as "long sleeve [MSK] [MSK] in black", rather than "long sleeve [MSK] #ie in [MSK]" occasionally. The input tokens are masked out with the probability of 15%. Given a text token sequence $t = \{t_1, t_2, …, t_n\}$, the masked-out sequence is denoted by $t_{\backslash i} = \{t_1, t_2, …, [MSK]_i, …, t_n\}$, which denotes token $i$ is masked out. The operator "\" means removing. The last-layer hidden outputs of the masked-out tokens are fed into a classifier over the standard BERT vocabularies. Finally, the MLM task is to minimize the cross-entropy loss, written as

$$l_{MLM}(\theta) = -E_{t \sim D} \log P(t_i | t_{\backslash i}, \theta) \quad (1)$$

where $\theta$ is the FashionBERT parameters and $D$ is the whole training set. $P(t_i | t_{\backslash i}, \theta)$ denotes the probability of the masked-out token $t_i$ predicted by FashionBERT, given surrounding tokens $t_{\backslash i}$.

Masked Patch Modeling (MPM): Similar to MLM, we mask out certain patches in a patch sequence in the MPM task. Given an image patch sequence $p = \{p_1, p_2, …, p_m\}$, we randomly mask out patches with the probability of 10%, denoting as $p_{\backslash i} = \{p_1, p_2, …, [MSK]_i, …, p_m\}$. The patch features of masked-out patches are set to zero. MPM is to predict the distribution over the masked-out patch features. The MPM training is supervised by minimizing the KL-divergence between the distributions of patch features.

$$l_{MPM}(\theta) = E_{KL_{p \sim D}}(Distr.(p_i | p_{\backslash i}, \theta) | Distr.(p_i)) \quad (2)$$

Text and Image Alignment (TIA): In the TIA task, the hidden output of the special token [CLS] is fed into a binary classifier to indicate whether the text and image data are matched. For one positive example in the train dataset, the text and image are extracted from one same fashion product, while for one negative sample, the text and image are randomly selected from different fashion products. TIA objects to optimize the binary cross-entropy loss.

$$l_{TIA}(\theta) = -E_{<t,p> \sim D}[y * \log P(\hat{y} | <t, p>, \theta) + (1 - y) * \log(1 - P(\hat{y} | <t, p>, \theta))] \quad (3)$$

where y and $\hat{y}$ denote the true and predicted labels, respectively.

In sum, FashionBERT attempts to optimize the aggregated loss function as seen in Equation (4), which is well acknowledged as an multitask learning problem.

$$\mathcal{L}(\theta) = \sum_{i=1}^{L} \omega_i l_i(\theta) \quad (4)$$

$L$ is the task number and in FashionBERT $L$ equals to three. ω are the loss weights to balance the learning of each task. We treat the determination of the loss weight $\omega_i$ as a new optimal problem:

$$\text{argmin} -\frac{1}{2}\sum_{i=1}^{L} \|\omega_i \nabla l_i\|^2 + \frac{1}{2}\sum_{i,j=1}^{L} \|\omega_i - \omega_j\|^2$$

$$s.t. \sum_{i=1}^{L} \omega_i = 1 \text{ and } \exists \omega_i \geq 0 \quad (5)$$

In Equation (5), on one hand we aim to minimum the total weighted loss, and on the other hand we expect FashionBERT fairly treats the learning of all tasks. Considering the KKT conditions (Karush-Kuhn-Tucher Conditions) [3], we can obtain the solution to $\omega_i$ as

$$\omega_i^* = \frac{(L - \nabla l_i^2)^{-1}}{\sum_{i=1}^{L}(L - \nabla l_i^2)^{-1}} \quad (6)$$

The proof to this solution is in the appendix. We illustrate the training procedure of FashionBERT in Algorithm.1.

---

**Algorithm 1** FashionBERT Training with Adaptive Loss

**Input**: the aggregated loss functions $\mathcal{L}(\theta) = \sum l_i(\theta)$ and training dataset $D$
**Parameter**: the model parameters $\theta$
**Output**: $\theta$
1:  Let $\omega_i = 1/L$.
2:  **for** each batch of train data **do**
3:      Feed the train batch into FashionBERT to get all $l_i$.
4:      Obtain weight losses according to Equation (6)
5:      Aggregate loss function according to Equation (4)
6:      Update FashionBERT by optimizing $\mathcal{L}(\theta)$ with ADAM $\theta = \theta - \eta \, \partial \mathcal{L}(\theta)/\partial \theta$
7:  **end for**
8:  **return** model parameters $\theta$

---

## 3. Experiments

In this section, we describe our experimental settings and show the main results.

### 3.1 Experimental Settings

| Evaluation | Approaches | Accuracy | Rank@1 | Rank@5 | Rank@10 |
|---|---|---|---|---|---|
| Image-to-Text | VSE | 63.14% | 4.01% | 11.03% | 22.14% |
| | VSE++ | 64.73% | 4.59% | 14.99% | 24.10% |
| | SCAN(LSE+AVG) | 65.04% | 4.59% | 16.50% | 26.60% |
| | PFAN | 72.01% | 4.29% | 14.90% | 24.20% |
| | ViLBERT–Zeroshot | 73.52% | 8.99% | 15.34% | 26.14% |
| | ViLBERT–Finetune | 88.11% | 20.97% | 40.49% | 48.21% |
| | VLBERT –Finetune | 85.42% | 19.26% | 39.90% | 46.05% |
| | **Our FashionBERT** | **91.01%** | **23.96%** | **46.31%** | **52.12%** |
| Text-to-Image | VSE | 60.28% | 4.35% | 12.76% | 20.91% |
| | VSE++ | 64.01% | 4.60% | 16.89% | 28.99% |
| | SCAN(LSE+AVG) | 65.74% | 4.30% | 13.00% | 22.30% |
| | PFAN | 72.66% | 6.20% | 20.79% | 31.52% |
| | ViLBERT–Zeroshot | 73.02% | 7.18% | 18.73% | 29.84% |
| | ViLBERT–Finetune | 88.61% | 21.12% | 37.23% | 50.11% |
| | VLBERT –Finetune | 85.51% | 22.63% | 36.48% | 48.52% |
| | **Our FashionBERT** | **91.09%** | **26.75%** | **46.48%** | **55.74%** |

Table 1: Comparison of FashionBERT with the baseline and SOTA pre-trained approaches.

**Datasets**: Although there exist several fashion datasets [4, 20, 21, 24, 37], the majority of these datasets only contain a limited number of images or lack necessary fashion descriptions. In our experiments, we adopt the Fashion-Gen dataset, which contains 67,666 fashion products. The Fashion products are associated with text descriptions provided by professional stylists, which are high-quality data for our FashionBERT pre-training. Each product is photographed from 1 to 6 different angles. We collect 293,008 image (256*256 pixels) and description pairs, among which 260,480 pairs are used for training, and 32,528 for testing. FashionBERT is pre-trained on the training dataset. We directly evaluate it on the testing dataset.

**Evaluation Tasks and Metrics**: We introduce two tasks (i.e., Text and Image Matching, and Cross-modal Retrieval) to test the FashionBERT performances.

For Text and Image Matching evaluation, all the test data are adopted. Given a text and image pair, the output of the TIA classifier is regarded as the matching similarity. We use Accuracy to assess the FashionBERT performance on this matching task.

For Cross-modal Retrieval evaluation, 1,000 unique description queries and 1,000 unique image queries are randomly selected from the test data. Given one description (or image) query, the ground-truth image (or description) from the same product and randomly-sampled 100 images (or description) from other products consist of the candidate rank set. Then we acquire the evaluation datasets for text-to-image and image-to-text retrieval. Same with [23], we score each description and image pair in the test dataset and then sort the candidate rank set. We use three metrics (i.e. Rank@K (K=1, 5, 10)) to evaluate FashionBERT on Cross-modal Retrieval. Rank@K is the percentage of ground-truth matchings appearing in the top-K ranked list.

**Implementation Details**: We reuse the pre-trained parameters from the 12-layer BERT-base. Each block has 768 hidden units and 12 self-attention heads. The text representation follows the same processing of BERT. The maximum sequence length is set to 512, among which the patch sequence is set to 64 (8*8 patches) and the maximum text sequence length is set to 448 (=512-64, including the special tokens). For each patch, ResNeXt101 [43] is first adopted to extract the patch features and the dimensions of patch features are 2048.

FashionBERT is implemented with Tensorflow[5] and trained on 8*Tesla V100 GPUs with early stopping to avoid overfitting (it usually had run about 10 epochs when stopping). In each training batch, 64 shuffled <Text, Image> pairs are utilized. Adam optimizer is applied with the learning rating of 2e-5, $\beta_1 = 0.95$, $\beta_2 = 0.999$, weight decay of 1e-4, learning rate warmed up at the first 5,000 steps, and then linear decay.

### 3.2 Evaluation of the SOTA and Pioneer Approaches

In this section, we conduct the experiments to response two questions:

- Does our model perform well comparing with the baseline approaches?
  We implement the following baseline approaches.
  **VSE** [11]: VSE directly projects image features and text features into visual semantic embedding space in an end-to-end manner, which is regarded as our first baseline approach.
  **VSE++** [10]: VSE++ extends the VSE approach with modification of the rank-based loss function which pays more attention to the hard negatives.
  **SCAN** [18]: SCAN performs the cross-modal match on the image RoIs and text tokens. Meanwhile, the attention mechanism [39] is leveraged to enhance matching ability.

---
[5] https://www.tensorflow.org/

**PFAN** [41]: The position information of each RoI is incorporated in the PFAN model. By this way, the matching can better understand the position information in both texts and images.
- Does our model perform well comparing with the state-of-the-art (SOTA) pre-trained approaches?
We compare our model with the pre-trained cross-modal models, ViLBERT[6] [23] and VLBERT[7] [34].
**ViLBERT-ZeroShot** [23]: In ViLBERT, the authors fine-tune and evaluate ViLBERT with cross-modal retrieval. They release the fine-tuned cross-modal retrieval model as well. Thus, in this experiment, we evaluate the Fashion-Gen testing data with the released ViLBERT model. We follow the same RoI extraction method in ViLBERT. For each image, we keep 10 to 36 high-scoring regions with Faster R-CNN [13] pre-trained on the Visual Genome dataset [27].
**ViLBERT-Finetune**: In this experiment, based on the pre-trained ViLBERT, we fine-tune a new cross-modal retrieval model with the Fashion-Gen training data.
**VLBERT-Finetune** [34]: The pre-trained VLBERT model is not evaluated with cross-modal retrieval. We thus fine-tune a new cross-modal retrieval model with the pre-trained VLBERT. In VLBERT, we follow its RoI extraction setting, where the minimum number of RoIs is set to 10, and at most 100 RoIs with detection scores higher than 0.5 are selected for each image.

All these approaches have the same settings, e.g., hidden layers, embedding size, patch features. Each experiment runs three times, and the average performances are shown in Table 1. we observe that FashionBERT with patch and adaptive loss achieves the significant improvement on Rank@K metrics (statistically significant difference over the other baselines with $p < 0.1$). This shows the excellent ability of FashionBERT in fashion text and image matching. In the ViLBERT-ZeroShot experiment, though ViLBERT is pre-trained with the general domain dataset, it does not perform well in matching of the fashion domain, which hints that there may exist a large gap between the fashion domain and the general domain. Thus, the pre-training model with general domain dataset contributes little to the matching of the fashion domain. This also explains that after finetuning with the Fashion-Gen dataset, a soar increase is achieved in the ViLBERT-Finetune experiment. Furthermore, it is found that FashionBERT benefits more from the patch method than from the RoI method compared with the RoI-based ViLBERT/VLBERT experiments and our Patch-based FashionBERT. As mentioned in section 1, more useless and repeated RoIs are detected from fashion images. The self-supervised learning of FashionBERT was expected to mask out some of these RoIs and to predict them with surrounding ones. However, surrounding tokens may either provide the irrelevant information (the useless RoIs, e.g., the head, hands of the models) or almost the same information (the repeated RoIs which generate the similar patch features as well). It is hard for the model to learn

[6] https://github.com/jiasenlu/vilbert_beta
[7] https://github.com/jackroos/VL-BERT

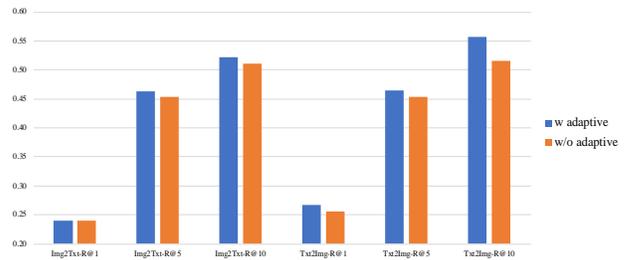
**Figure 3: FashionBERT with and without Adaptive Loss**

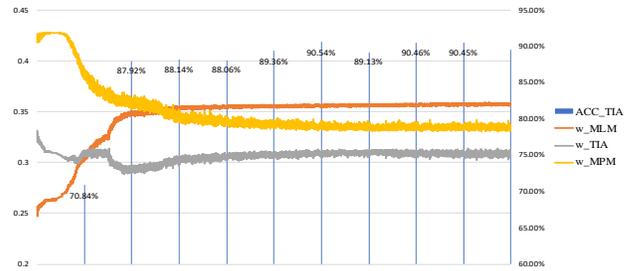
**Figure 4: average $\omega_i$ during the training batches**

| Metric | Accuracy | Rank@1 | Rank@5 | Rank@10 |
|---|---|---|---|---|
| V3-Img2Txt | 88.84% | 20.43% | 42.97% | 50.01% |
| V3-Txt2Img | 89.81% | 24.31% | 45.49% | 50.27% |
| RNX-Img2Txt | 91.01% | 23.96% | 46.31% | 52.12% |
| RNX-Txt2Img | 91.09% | 26.75% | 46.48% | 55.74% |

**Table 2: Evaluation of patch feature extraction, where "V3" and "RNX" refer to the pre-trained inceptionV3 and ResNeXt-101models**

useful information from these noise regions. On the contrary, the patches provide non-repeated and reasonably-related information, which is more suitable for self-supervised learning and enhances the performance of the fashion text and image matching.

### 3.3 Ablation Studies
In this section, we conduct ablation experiments in order to incrementally exam the influences of adaptive loss, pre-trained image models and model size of BERT.

**Effect of Adaptive Loss**: We first investigate the contribution of the task-agnostic adaptive loss algorithm. We test FashionBERT with and without adaptive loss. When without adaptive loss, $\omega_i$ is set to $1/L$.

The evaluation results are showed in Figure 3, which illustrates the adaptive loss weight produces positive influence on the FashionBERT performance. As shown in Equation (6), $\omega_i$ is in direct ratio to $l_i$. This means that when the task $i$ gets a larger loss, the adaptive loss weight $\omega_i$ will be larger than the others. In consequence, FashionBERT will pay more attention to the learning of the task $i$. We show the adaptive loss weights during the training batches in Figure 4. It is found that at the beginning FashionBERT pays more attention to MLM and TIA since these two tasks are newly introduced in BERT. Later on, the weight of

TIA and MPM decay. At the same time, we find that when FashionBERT well matches the fashion texts and images (as seen ACC_TIA in Figure 4), it shifts its attention on the MPM and MLM tasks. These two tasks are relatively harder than the TIA task. This may also hint that there is still room for further improvements if more difficult matching tasks can be introduced in, for example, the token-level and patch-level alignment.

**Effect of pre-trained image models**: We compare different pre-trained image models when extracting patch features. In this section, we compare ResNeXt-101 with InceptionV3. For the sake of fair comparison, we use their pre-trained parameters in InceptionV3 and ResNeXt-101, respectively. The dimensions of the patch features are set to 2048 and the same hyperparameters are used in FashionBERT.

The evaluation results are illustrated in Table 2, where FashionBERT with ResNeXt-101 shows clearly better results than that with InceptionV3. Compared with InceptionV3, ResNeXt-101 contains a series of residual network blocks. This residual structure brings more raw-pixel information into the Transformer encoders, which helps the modeling. We will explore more pre-trained tasks and extract more representative information.

**Effect of Model Size**: In order to test the influence of the model size, we vary the number of the transformer encoder layers. FashionBERT is experimented with 4-layer, 6-layer and 12-layer transformer encoders. The 4-layer FashionBERT means that we load the first four layers of the pre-trained BERT. The other hyperparameters are consistent among all the experiments.

The evaluation results are shown in Table 3, which demonstrates that FashionBERT benefits from the deeper BERT encoders over all the metrices except Rank@1 (Rank@1 tends to be more sensitive than the other two). Due to limited resources, we did not experiment with the pre-trained BERT-Large. It is possible to achieve further improvements with the 24-layer BERT model.

| Metric | Accuracy | Rank@1 | Rank@5 | Rank@10 |
|---|---|---|---|---|
| 4L-Img2Txt | 82.78% | 20.11% | 43.88% | 49.77% |
| 4L-Txt2Img | 84.50% | 25.43% | 44.15% | 50.92% |
| 6L-Img2Txt | 88.52% | 21.23% | 44.03% | 50.23% |
| 6L-Txt2Img | 89.93% | 26.91% | 45.81% | 54.17% |
| 12L-Img2Txt | 91.01% | 23.96% | 46.31% | 52.12% |
| 12L-Txt2Img | 91.09% | 26.75% | 46.48% | 55.74% |

**Table 3: Evaluation of model size, where L denotes Layer**

## 3.4 Industry Applications

FashionBERT is a general text and image encoder in the fashion domain, which can be widely applied in varieties of text and image matching tasks. The vanilla application is end-to-end cross-modal retrieval in practice, where we vectorize the search queries and the products with FashionBERT and then perform retrieval and ranking with nearest-neighbor search [15, 45]. The main framework of cross-modal retrieval is shown in Figure 5.

In practice cross-modal retrieval, FashionBERT is pretrained and fine-tuned with private datasets from scratch. The pre-train

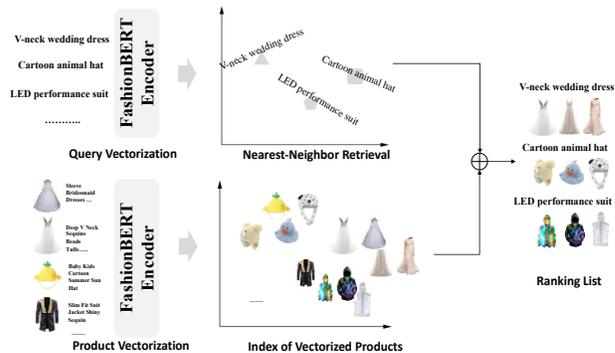

**Figure 5: FashionBERT application in cross-modal retrieval**

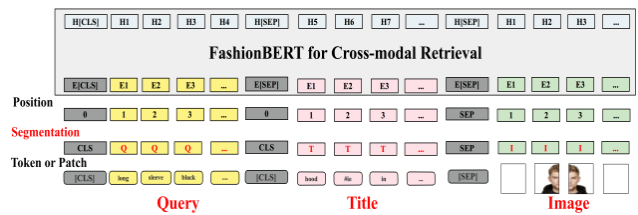

**Figure 6: Fine-tune model of cross-modal retrieval for <Query, Title, Image> triple inputs**

<Text, Image> pairs are collected from fashion products in our Alibaba.com[8] website, where the titles of products act as the text information. The fine-tune dataset of cross-modal retrieval is extracted from logs of the search engine. From search logs, the queries and their clicked products first compose of the click dataset. In consequence, <Query, Title, Image> triples are chosen as the fine-tune dataset, where "Title" and "Image" are from the same clicked products. Totally, ten million <Title, Image> pairs are collected for pre-training. Two million <Query, Title, Image> triples are collected for fine-tuning, of which ninety percent are used in training and the rest are used in testing. Like the downstream task of Visual Question Answer (VQA) in VLBERT [34], "Q", "T" and "I" are utilized in the segmentation to distinguish the three input types of "Query", "Title" and "Image" in fine-tuning. We show the fine-tune model of cross-modal retrieval in Figure 6.

The previous experiments already prove the advantages of FashionBERT in the public dataset. In this set of experiments, we focus on the comparison of cross-modal retrieval and single-modal retrieval. To certain degree, the fine-tune task can be regarded as the click prediction in information retrieval. Thus, we employ the accuracy and AUC metrics to evaluate the performance.

---

[8] www.alibaba.com - the global wholesale online trading market

**oriBERT**: This approach is the online baseline approach. The image information is not incorporated either in pre-training or in fine-tuning. In pre-training, we adopt the original pre-trained BERT model [8]. In fine-tuning, <Query, Title> pairs from <Query, Title, Image> triples compose of the training and testing datasets. Thus, this approach can be regarded as the single-model retrieval approach.

**BERT+IMG**: In this experiment, image information is not interacted with text information in BERT. Rather than feeding image features into BERT, we concatenate them with the BERT outputs of the final layer for click prediction in fine-tuning.

Furthermore, the inference speed is pivotal for online deployment. The computational and memory complexity of self-attention in BERT is $O(K^2D)$, highly related to the input sequence length $K$ ($K = m + n$ in our FashionBERT) and the hidden dimension $D$ [39]. The performance can be improved with a shorter sequence length. To speed up the online inference, we attempt the Variable Sequence Length (VSL) strategy, which directly concatenates the text and patch sequences, rather than first pads either of them to the max length.

The evaluation results are shown in Table 4. First, both the BERT+IMG and FashionBERT approaches benefit a lot from image features compared with oriBERT. Meanwhile, image features further enhance the performances after fully interacting with text features in FashionBERT. Second, though better performances are obtained with six-layer BERT or FashionBERT, the inference latencies vastly excel the online requirements. To trade off inference speeds and matching performances, model reduction has to be introduced in and only the first two FashionBERT layer in our final model. Meanwhile, we observe that the VSL strategy is quite effective in accelerating the speed and has very little influence on performances.

Online inference is still one of the most challenging issues for BERT-like models in deployment. More recently, some tiny varieties of BERT are proposed to address this issue, such as tinyBERT [16], ALBERT [17], AdaBERT [5], Reformer [26]. We are now attempting to incorporate these approaches in FashionBERT to further accelerate the online performance.

## 4. Related Work

### 4.1 Pre-training

The pre-training technique recently has been widely adopted in Machine Learning, which allows the learning model to leverage information from other related tasks.

The pre-training technique becomes popular first in CV. Krizhevsky *et al*. propose AlexNet in 2012 [28], with which they win the 2012 ILSVR image classification competition [7]. Later on, the researchers found that these CNN blocks in AlexNet pre-trained on ImageNet or the other large-scale image corpus can be treated as general feature extractors and perform well in various of downstream tasks [9]. Since then the researchers propose more effective CNN-based models and pre-train them on massive dataset, such as VGG [33], Google Inception [36], ResNet [43].

| Metric | Accuracy | AUC | Latency(ms) |
|---|---|---|---|
| 6L-BERT | 71.21% | 0.8121 | 51.5 |
| 6L-BERT+IMG | 74.42% | 0.8283 | 62.4 |
| **6L-FashionBERT** | **75.21%** | **0.8387** | **66.9** |
| 2L-BERT | 67.81% | 0.7746 | 17.2 |
| 2L-BERT+IMG | 70.31% | 0.7836 | 20.8 |
| 2L-FashionBERT | 72.47% | 0.8018 | 22.3 |
| **2L-FashionBERT(VSL)** | **72.43%** | **0.8009** | **18.3** |

**Table 4: Evaluation of FashionBERT in fine-tuning. All approaches are tested on the Intel(R) Xeon(R) E5-2650 servers.**

These pre-trained models are widely adopted in the CV tasks, like object detection [12] and semantic segmentation [22]. The applications of the pre-training technique in NLP is behind in CV. The early studies on pre-training in NLP can be traced back to word embedding, such as word2vec, GloVe [30]. Transformer is proposed to leverage the self-attention mechanism to draw global dependencies between inputs and outputs [39]. After that, a series of extended studies are presented for example, GPT [31], BERT [8], XLNet [46], among which BERT is one of the most popular models for its performances in variants of NLP tasks. Very recently, the self-supervised learning draws increasing attention of researchers, which allow the model to pre-train with large-scale unlabeled data and learn a more general representation across tasks.

### 4.2 Text and Image Matching

There is a long research history on text and image matching. These researches have greatly promoted the developments of the cross-modal applications, such as cross-modal retrieval [40], image captioning [1] and visual question answering [2].

Hardoon *et al*., overview the applications of Canonical Correlation Analysis (CCA), where CCA is introduced to project text and image features into a shared vector space by maximizing the cross relation [14]. Researchers then propose variants of CCA-based approaches [25, 32, 44]. Visual-semantic embedding (VSE) present by Frome *et al*., learns semantic relationships between labels and explicitly maps images into the semantic embedding space [11]. Faghri *et al*., extend VSE and propose VSE++ by introducing the hard-negative mining technique [10]. SCAN performs the cross-modal match on the image RoIs and text tokens. Meanwhile, the attention mechanism is leveraged to enhance matching ability [18]. The position information of each RoI is incorporated in the PFAN model [41]. By this way, the matching can better understand the position information in both texts and images. Recently, with the success of pre-training and self-supervised learning [29, 43], researchers attempt to apply BERT in cross-modal tasks. VideoBERT [35] leverages off-the-shelf networks for action recognition from video clips. The clusters of video clips are regarded as visual words. VideoBERT learns the high-level video representation by using BERT as the backbone network. ViLBERT [23] focuses on the representation learning of the general domain. It extracts RoIs from images and regards these RoIs as image tokens. Unicoder-VL [19] and VL-BERT [34] follow the same RoI method, but select a single cross-modal Transformer structure to allow text and image information interacting earlier.

In this paper, we mainly follow this research line. We argue that this RoI method used in these BERT-based approaches does not work well in the fashion domain since the detected RoIs from fashion images are not fine-grained enough for fashion text and image matching.

## 5. Conclusions

In this paper, we focus on the text and image matching in cross-modal retrieval of the fashion domain. We propose FashionBERT to address the matching issues in the fashion domain. FashionBERT splits images into patches. The images patches and the text tokens are as the inputs of the BERT backbone. To trade off the learning of each task, we present the adaptive loss algorithm which automatically determines the loss weights. Two tasks are incorporated to evaluate FashionBERT and extensive experiments are conducted on the Fashion-Gen dataset. The main conclusions are 1) the patch method shows its advantages in matching fashion texts and images, compared with the object-level RoI method; 2) through the adaptive loss, FashionBERT shifts its attention on different tasks during the training procedure.

Compared with the matching of the general domain, there is still room for further improvements in the fashion domain. In the future, 1) To better understand the semantic of the fashion images, we attempt to construct more fine-grained training task (for example, token-level and patch-level alignment) to force FashionBERT to learn more detail information. 2) We attempt to visualize the FashionBERT matching secrets. This would help to understand how FashionBERT work inside and make further improvement. 3) We are attempting the model reduction, knowledge distillation approaches to further speed up the online inference.

## APPENDIX

**PROOF**. The adaptive loss weights learning can be written as

$$\operatorname*{argmin}_{\omega^*} -\frac{1}{2}\sum_{i=1}^{L}\|\omega_i \nabla l_i\|^2 + \frac{1}{2}\sum_{i,j=1}^{L}\|\omega_i - \omega_j\|^2$$
$$s.t. \sum_{i=1}^{L} \omega_i = 1 \text{ and } \exists \omega_i \geq 0 \quad (1)$$

We first omit the non-negative constraint and apply the Lagrange multipliers and get the Lagrange:

$$L(\omega, \alpha) = -\frac{1}{2}\sum_{i=1}^{L}\|\omega_i \nabla l_i\|^2 + \frac{1}{2}\sum_{i,j=1}^{L}\|\omega_i - \omega_j\|$$
$$+ \alpha(1 - \sum_{i=1}^{L} \omega_i) \quad (2)$$

The solution is obtained by

$$\nabla_\omega L(\omega, \alpha) = 0 \quad (3.1)$$

$$\nabla_\alpha L(\omega, \alpha) = 0 \quad (3.2)$$

From Equation (3.1), we get

$$-\nabla l_i^2 \omega_i + \sum_{i,j=1}^{L} \omega_i - \sum_{i,j=1}^{L} \omega_j - \alpha = 0$$
$$\Rightarrow \omega_i = \frac{1+\alpha}{L - \nabla l_i^2} \quad (4)$$

From Equation (3.2), we get

$$\sum_{i=1}^{L} \omega_i = 1 \quad (5)$$

By taking Equation (4) into Equation (5), the solution can be achieved by

$$\omega_i^* = \frac{(L - \nabla l_i^2)^{-1}}{\sum_{i=1}^{L}(L - \nabla l_i^2)^{-1}} \quad (6)$$

when $\nabla l_i \in [0,1)$, $\omega_i^*$ is non-negative and satisfy the non-negative constrain in Equation (1).

## ACKNOWLEDGMENTS